\documentclass[prl,twocolumn,showpacs,preprintnumbers,amsmath,amssymb]{revtex4}

\usepackage{graphicx}% Include figure files
\usepackage{dcolumn}% Align table columns on decimal point

\begin{document}

%\preprint{APS/123-QED}

\title{Third and Fourth Order Phase Transitions:\\
Exact Solution for the Ising Model on the Cayley Tree}

\author{Borko D. Sto\v si\' c}
\email{borko@ufpe.br}
\affiliation{
Departamento de Estat\' \i stica e Inform\' atica, 
Universidade Federal Rural de Pernambuco,\\
Rua Dom Manoel de Medeiros s/n, Dois Irm\~ aos,
52171-900 Recife-PE, Brasil
}
\author{Tatijana Sto\v si\' c}
\affiliation{
Laboratory for Theoretical Physics, 
Institute for Nuclear Sciences, \\
Vin\v ca, P.O. Box 522, YU-11001 Belgrade, Yugoslavia
}
\author{Ivon P. Fittipaldi}
\affiliation{
Minist\' erio da Ci\^ encia e Tecnologia, 
Esplanada dos Minist\'erios, Bloco E, 2$^o$ andar, Sala 215, 70067-900 Bras\' \i lia-DF, Brazil
}

\date{\today}% It is always \today, today,

\begin{abstract}
An exact analytical derivation is presented, showing that the Ising model 
on the Cayley tree exhibits a line of third order phase transition points, 
between temperatures $T_2=2k_B^{-1}J\ln ({\sqrt 2}+1) $
and $T_{BP}=k_B^{-1}J\ln (3)$, and a line of fourth order phase
transitions between $T_{BP}$ and $\infty$, where $k_B$ is the Boltzmann 
constant, and $J$ is the nearest-neighbor interaction parameter.
\end{abstract}

% PACS, the Physics and Astronomy Classification Scheme.
%05.50.+q	Lattice theory and statistics (Ising, Potts, etc.)
%64.60.Cn	Order-disorder transformations and statistical mechanics of model systems
%75.10.Hk	Classical spin models

%\keywords{Cayley tree, Ising model, Phase transitions}%Use showkeys class option if keyword
\maketitle

Ever since the pioneering works on phase transitions and critical phenomena, 
it has been clear that phase transitions of higher order are conceptually 
possible, but, to the best of our knowledge, 
up to date there has been no rigorous proof of existence of a single 
system exhibiting a phase transition of finite order higher then two. 

One notable attempt in this direction is the 
work of M\" uller-Hartmann and Zittartz \cite{muller}, interpreting 
the series expansion of the free energy of the Ising model on the Cayley tree
in the limit of small field $H\longrightarrow 0^+$, in terms of phase 
transitions of all even orders (from two to infinity).
More precisely, their claim was that the susceptibility of order $2\ell$ 
(i.e. the 2$\ell$-th derivative of the free energy with respect to field), 
diverges at temperature
$T_{2\ell}=k_B^{-1}J\ln [1+2/(B^{1-1/2\ell}-1)]$ for $\ell=1,2,\dots ,\infty$, 
where $B$ is the tree branching number (coordination number minus 1). Thus, the
zero field susceptibility was found to diverge from $T=0$ to 
$T_2=k_B^{-1}J\ln [({\sqrt B}+1)/({\sqrt B}-1)]$, representing a line of 
second order phase transitions (with respect to field), fourth order susceptibility
was found to diverge between $T_2$ and
$T_4=k_B^{-1}J\ln [(B^{3/4}+1)/(B^{3/4}-1)]$, representing a line of 
fourth order phase transitions, while the infinite order phase transition 
was found to occur at the Bethe-Pierls temperature
$T_{BP}=k_B^{-1}J\ln \left[(B+1)/(B-1)\right]$.

It is shown in this work through an exact analytical calculation of the recursion
relations for the derivatives
of the free energy, that the exact expressions contain terms previously unnoticed 
 \cite{muller} in the approximate analysis. 
Interpratation of these results requires some special consideration of symmetry
breaking, the classical approach not being adequate for the system at hand.
It is found that the third order derivative
diverges between $T_2$ and $T_{BP}$ (all the lower derivatives being finite), 
representing a line of third order phase
transitions, while the fourth order derivative diverges between $T_{BP}$ and
$\infty$ (where all lower derivatives are finite), 
representing a fourth order phase transition line.

We consider the nearest-neighbor Ising model with the Hamiltonian
\begin{equation}
{\cal H}=-J\sum_{\langle nn\rangle}S_iS_j-H\sum_{i}S_i,\label{one}
\end{equation}
where $J$ is the coupling constant, $H$ is the external magnetic field,
$S_i=\pm 1$ is the spin at site $i$, and $\langle nn\rangle$ denotes
summation over the nearest-neighbor pairs.  
For simplicity, hereafter we consider only the tree with $B=2$,
while the analysis which follows may be directly generalized for 
arbitrary tree branching number.
Following Eggarter
\cite{egarter}, we further consider systems situated on a single
$n$-generation branch of a Cayley tree, composed of two 
$(n-1)$-generation branches connected to a single initial site. 
Thus, the
$n$-generation branch consists of $N_n=2^{n+1}-1$ spins, the
$0$-generation branch being a single spin.  
The exact recursion relations for
the partial partition functions of any two consecutive generation
branches are easily derived \cite{egarter} to be
\begin{equation}
Z_{n+1}^{\pm}=y^{\pm 1}\left[ x^{\pm 1}Z_{n}^{+} 
+ x^{\mp 1}Z_{n}^{-}\right] ^2,\label{two}
\end{equation}
where $Z_{n}^{+}$ and $Z_{n}^{-}$ denote the partition functions
restricted by fixing the initial spin (connecting the two
$n$-generation branches) into the $\{ +\}$ and $\{ -\}$ position,
respectively, and where we have used the notation $x\equiv exp(\beta
J)$ and $y\equiv exp(\beta H)$, with $\beta=1/k_BT$ denoting the
reciprocal of the product of the Boltzmann constant $k_B$ and the
temperature $T$.
The usual approach of attempting to establish the
field dependent expression for the partition function in the
thermodynamic limit, then finding its field derivatives, and finally
taking the limit $H\longrightarrow 0$ yields only approximate
solutions for the zero-field susceptibility
\cite{muller,heimburg,matsuda}. 
In an earlier work \cite{SSF}, current authors have derived an exact 
expression for the zero-field magnetization and susceptibility, 
by using
the strategy of finding the recursion relations for the {\it
field derivatives} of the partition function, taking the limit
$H\longrightarrow 0$, and only then performing the actual iterations to
reach the thermodynamic limit.
Here we extend this approach to find higher derivatives
of the partition function.

Equation (\ref{two}) can be formally differentiated with respect to field, to
find the recursion relations for the field derivatives of the partition function.
Up to the forth derivative 
we find the following recursion relations
\begin{eqnarray}
&&{\frac{\partial Z_{n+1}^{\pm}}{\partial h}}=
y^{\pm 1} \left (
\pm {\Gamma_n^{0\pm}}^{2}+2 {\Gamma_n^{0\pm}}{\Gamma_n^{1\pm}}
 \right ), 
\label{three}
\end{eqnarray}
\begin{eqnarray}
{\frac{\partial^2 Z_{n+1}^{\pm}}{\partial h^2}}=&&
y^{\pm 1} \left [
{\left(\Gamma_n^{0\pm}\right)}^{2} 
\pm 4{\Gamma_n^{0\pm}}{\Gamma_n^{1\pm}}  
+2 {\left(\Gamma_n^{1\pm}\right)}^2 +\right.\nonumber\\
&& \left.+2 {\Gamma_n^{0\pm}}{\Gamma_n^{2\pm}} 
\right] ,
\label{four}
\end{eqnarray}
\begin{eqnarray}
{\frac{\partial^3 Z_{n+1}^{\pm}}{\partial h^3}}=&&
y^{\pm 1} \left [
\pm {\left(\Gamma_n^{0\pm}\right)}^{2} +
6 {\Gamma_n^{0\pm}}{\Gamma_n^{1\pm}}  \pm 
6 {\left(\Gamma_n^{1\pm}\right)}^2 \pm \right.\nonumber\\
&& \left. \pm 6 {\Gamma_n^{0\pm}}{\Gamma_n^{2\pm}} +
6 {\Gamma_n^{1\pm}}{\Gamma_n^{2\pm}} +
2 {\Gamma_n^{0\pm}}{\Gamma_n^{3\pm}}
\right ] ,
\label{five}
\end{eqnarray}
\begin{eqnarray}
{\frac{\partial^4 Z_{n+1}^{\pm}}{\partial h^4}}=&&
y^{\pm 1} \left [ 
{\left(\Gamma_n^{0\pm}\right)}^{2} \pm
8 {\Gamma_n^{0\pm}}{\Gamma_n^{1\pm}} +
12 {\left(\Gamma_n^{1\pm}\right)}^2 + \right.\nonumber\\
&& +12 {\Gamma_n^{0\pm}}{\Gamma_n^{2\pm}}  \pm
24 {\Gamma_n^{1\pm}}{\Gamma_n^{2\pm}} \pm
8 {\Gamma_n^{0\pm}}{\Gamma_n^{3\pm}} + \nonumber\\ 
&& \left. +6{\left(\Gamma_n^{2\pm}\right)}^2 + 
8 {\Gamma_n^{1\pm}}{\Gamma_n^{3\pm}} +
2 {\Gamma_n^{0\pm}}{\Gamma_n^{4\pm}}\right ] ,
\label{six}
\end{eqnarray}
where we have used notation $h\equiv \beta H$ and
\begin{eqnarray*}
{\Gamma_n^{i\pm}}\equiv
{\frac{\partial^i Z_{n}^{+}}{\partial h^i}}{x^{\pm 1}}+
{\frac{\partial^i Z_{n}^{-}}{\partial h^i}}{x^{\mp 1}}.
\end{eqnarray*}

Starting from a single spin (0-th generation branch), for which we have
${{\partial^k Z_{0}^{\pm}}/{\partial h^k}}=(\pm 1)^k y^{\pm 1}$,
it is straightforward to show by mathematical induction, using
(\ref{two}-\ref{six}),
that for {\it zero field}
($y=1$) the symmetry equations
\begin{eqnarray}
%&&Z_n^{+}=Z_n^{-},\nonumber\\
%&&{\frac{\partial Z_{n}^{+}}{\partial \beta H}}=
%-{\frac{\partial Z_{n}^{-}}{\partial \beta H}},\nonumber\\
%&&{\frac{\partial^2 Z_{n}^{+}}{\partial (\beta H^2)}}=
%{\frac{\partial^2 Z_{n}^{-}}{\partial (\beta H)^2}},\nonumber\\
%&&{\frac{\partial^3 Z_{n}^{+}}{\partial (\beta H^3)}}=
%-{\frac{\partial^3 Z_{n}^{-}}{\partial (\beta H)^3}},\nonumber\\
%&&{\frac{\partial^4 Z_{n}^{+}}{\partial (\beta H^4)}}=
%{\frac{\partial^4 Z_{n}^{-}}{\partial (\beta H)^4}},
\left.{\frac{\partial^k Z_{n}^{+}}{\partial (\beta H)^k}}\right|_{H=0}=
\left.(-1)^k{\frac{\partial^k Z_{n}^{-}}{\partial (\beta H)^k}}\right|_{H=0},
\label{seven}
\end{eqnarray}
hold for a branch of arbitrary generation n. 
It then follows from (\ref{two}-\ref{seven})
that the moments
${\cal S}_n\equiv \frac{1}{Z_n^+}{\frac{\partial Z_{n}^{+}}{\partial h}}$, 
${\cal T}_n\equiv \frac{1}{Z_n^+}{\frac{\partial^2 Z_{n}^{+}}{\partial h^2}}$, 
${\cal U}_n\equiv \frac{1}{Z_n^+}{\frac{\partial^3 Z_{n}^{+}}{\partial h^3}}$, and
${\cal V}_n\equiv \frac{1}{Z_n^+}{\frac{\partial^4 Z_{n}^{+}}{\partial h^4}}$
satisfy the recursion relations,
\begin{eqnarray}
&{\cal S}_{n+1}=1+2\;t\;{\cal S}_n , \;\;\;\;\;\;\;\;\;\;\;\;\;\;\;\;\;\;\;\;& {\cal S}_0=1,\nonumber\\
&{\cal T}_{n+1}=1+4\;t\;{\cal S}_n+2\;t^2\;{\cal S}_n^2+2{\cal T}_n, \;\;\;\;\;\;\;\;\;\;&{\cal T}_0=1,\nonumber\\
&{\cal U}_{n+1}=1+6\;t\;{\cal S}_n+6\;t^2\;{\cal S}_n^2+6{\cal T}_n +\;\;\;\;\;\;\;\;\;\;&\nonumber\\
&+6\;t\;{\cal S}_n\;{\cal T}_n+2\;t\;{\cal U}_n &{\cal U}_0=1,\nonumber\\
&{\cal V}_{n+1}=1+8\;t\;{\cal S}_n+12\;t^2\;{\cal S}_n^2+12{\cal T}_n
\;\;\;\;\;\;\;\;\;\;&\nonumber\\
&+24\;t\;{\cal S}_n\;{\cal T}_n+8\;t\;{\cal U}_n\;\;\;\;\;\;\;\;\;\;&\nonumber\\
&+6\;{\cal T}_n^2+8\;t^2\;{\cal S}_n\;{\cal U}_n +2\;{\cal V}_n\;\;\;\;\;\;\;\;\;\;&{\cal V}_0=1,\nonumber\\
\label{eight}
\end{eqnarray}
where $t\equiv \tanh(\beta J)$.
Relations (\ref{eight}) can be iterated (by summing the
geometric series) to yield closed-form expressions for ${\cal S}_n$, 
${\cal T}_n$, ${\cal U}_n$ and ${\cal V}_n$ for arbitrary tree generation $n$,
and we find
\begin{eqnarray}
{\cal S}_n=
{\frac {{2}^{n+1}{t}^{n+1}-1}{2\,t-1}}
\label{nine}
\end{eqnarray}

\begin{widetext}
\begin{eqnarray}
{\cal T}_n=&&
{\frac {4\,{t}^{4}\left (4\,{t}^{2}\right )^{n}}
{\left (2\,{t}^{2}-1\right )\left (2\,t-1\right )^{2}}}
-{\frac {{2}^{n+1}\left (t+1\right )^{2}}{2\,{t}^{2}-1}}
+{\frac {4\,{t}^{2}\left (2\,t\right )^{n}}{\left (2\,t-1\right )^{2}}
+{\frac {2\,{t}^{2}-1}{\left (2\,t-1\right )^{2}}}}
\label{ten}
%                                  1
%{\frac {-1+2\,{t}^{2}}{\left (-1+2\,t\right )^{2}}}
%                                  2
%4\,{\frac {{t}^{4}\left (4\,{t}^{2}\right )^{n}}{\left (-1+2\,{t}^{2}
%\right )\left (-1+2\,t\right )^{2}}}
%                                  3
%-2\,{\frac {{2}^{n}\left (t+1\right )^{2}}{-1+2\,{t}^{2}}}
%                                  4
%4\,{\frac {{t}^{2}\left (2\,t\right )^{n}}{\left (-1+2\,t\right )^{2}}}
\end{eqnarray}
%\end{widetext}
%\begin{widetext}
\begin{eqnarray}
{\cal U}_n=&&
+{\frac {24\,{t}^{5}\left (8\,{t}^{3}\right )^{n}}{\left (2\,t+1\right )\left (
2\,{t}^{2}-1\right )\left (2\,t-1\right )^{4}}}
-{\frac {12\,\left (t+1\right )^{2}t\left (4\,t\right )^{n}}{\left (2\,t -1
\right )\left (2\,{t}^{2}-1\right )}}
+{\frac {12\,{t}^{3}\left (4\,{t}^{3}+2\,{t}^{2}-t-2\right )\left (4\,{t}^{2}
\right )^{n}}{\left (2\,{t}^{2}-1\right )\left (2\,t-1\right )^{4}}}\nonumber\\
&&+{\frac {\left (2\,t\right )^{n+1}\left (24\,n{t}^{4}-30\,n{t}^{2}+6\,n
+40\,{t}^{4}-16\,{t}^{3}-60\,{t}^{2}-2\,
t+11\right )}{\left (2\,t-1\right )^{4}\left (2\,
t+1\right )}}
+{\frac {6\,{2}^{n}\left (t+1\right )^{2}}{\left (2\,t-1\right )\left (2\,{t}^
{2}-1\right )}}\nonumber\\
&&-{\frac {8\,{t}^{3}-6\,{t}^{2}-6\,t+5}{\left (2\,t-1\right )^{4}}}
\label{eleven}
%                                  1
%12\,{\frac {{t}^{3}\left (4\,{t}^{3}+2\,{t}^{2}-t-2\right )\left (4\,{
%t}^{2}\right )^{n}}{\left (-1+2\,{t}^{2}\right )\left (-1+2\,t\right )
%^{4}}}
%                                  2
%6\,{\frac {{2}^{n}\left (t+1\right )^{2}}{\left (-1+2\,t\right )\left 
%(-1+2\,{t}^{2}\right )}}
%                                  3
%24\,{\frac {{t}^{5}\left (8\,{t}^{3}\right )^{n}}{\left (2\,t+1\right 
%)\left (-1+2\,{t}^{2}\right )\left (-1+2\,t\right )^{4}}}
%                                  4
%-12\,{\frac {\left (t+1\right )^{2}t\left (4\,t\right )^{n}}{\left (-1
%+2\,t\right )\left (-1+2\,{t}^{2}\right )}}
%                                  5
%2\,{\frac {\left (2\,t\right )^{n}t\left (40\,{t}^{4}-16\,{t}^{3}-60\,
%{t}^{2}-2\,t+11+24\,{t}^{4}n-30\,{t}^{2}n+6\,n\right )}{\left (2\,t+1
%\right )\left (-1+2\,t\right )^{4}}}
%                                  6
%-{\frac {8\,{t}^{3}-6\,{t}^{2}-6\,t+5}{\left (-1+2\,t\right )^{4}}}
\end{eqnarray}
\end{widetext}

\begin{widetext}
\begin{eqnarray}
{\cal V}_n=&&
{\frac {48\,{t}^{8}\left (12\,{t}^{2}-5\right )\left (16\,{t}^{4}\right )^{n}}{
\left (2\,t+1\right )\left (8\,{t}^{4}-1\right )\left (2\,{t}^{2}-1\right )^{2}
\left (2\,t-1\right )^{5}}}
+{\frac {12\,\left (t+1\right )^{4}{4}^{n}}{\left (2\,{t}^{2}-1\right )^{2}}}
-{\frac {48\,{t}^{4}\left (t+1\right )^{2}\left (8\,{t}^{2}\right )^{n}}
{\left (2\,t-1\right )^{2}\left (2\,{t}^{2}-1\right )^{2}}}
\nonumber\\
&&+{\frac {96\,{t}^{6}\left (8\,{t}^{4}+16\,{t}^{3}-6\,t-3\right )\left (8\,{t}^{3}\right )^{n}}
{\left (2\,{t}^{2}-1\right )\left (2\,t-1\right )^{5}
\left (2\,t+1\right )\left (4\,{t}^{3}-1\right )}}
-{\frac {48\,\left (t+1\right )^{2}{t}^{2}\left (4\,t\right )^{n}}
{\left (2\,{t}^{2}-1\right )\left (2\,t-1\right )^{2}}}
\nonumber\\
&&+{\frac {8\,{t}^{4}\left (64\,{t}^{6}+80\,{t}^{5}-104\,{t}^{4}-120\,{t}^{3}+50\,{t
}^{2}+40\,t-1\right )\left (4\,{t}^{2}\right )^{n}}
{\left (2\,t-1\right )^{5}\left (2\,t+1\right )\left (2\,{t}^{2}-1\right )^{2}}}
+{\frac {96\,\left (t^2-1\right ){t}^{4}\left (n+1\right )\left (
4\,{t}^{2}\right )^{n}}{\left (2\,{t}^{2}-1\right )\left (2\,t-1\right )^{4}}}
\nonumber\\
&&+{8\,\frac {\left (8\,{t}^{4}-20\,{t}^{3}-30\,{t}^{2}+2\,t+7\right ){t}^{2}
\left (2\,t\right )^{n}}{\left (2\,t-1\right )^{5}\left (2\,t+1\right )}}
+{\frac {48\,{t}^{2}\left (t^2-1\right )\left (n+1\right )\left (2
\,t\right )^{n}}{\left (2\,t-1\right )^{4}}}
\nonumber\\
&&-{\frac {{2}^{n+2}\left (t+1\right )\left (320\,{t}^{12}-784\,{t}^{10}-384\,{t}^
{9}+176\,{t}^{8}+372\,{t}^{7}+192\,{t}^{6}-56\,{t}^{4}-45\,{t}^{3}-11\,{t}^{2}+6
\,t+4\right )}{\left (2\,{t}^{2}-1\right )^{2}\left (8\,{t}^{4}-1\right )\left (
2\,t-1\right )^{2}\left (4\,{t}^{3}-1\right )}}
\nonumber\\
&&+{\frac {16\,{t}^{5}-24\,{t}^{4}-8\,{t}^{3}+28\,{t}^{2}-6\,t-5}
{\left (2\,t-1\right )^{5}}}
\label{twelve}
\end{eqnarray}
\end{widetext}

Before proceeding with the analysis of the above exact expressions, a word is
due on symmetry breaking. It has been commonly accepted for this model that all odd
derivatives of the free energy with respect to field are identically zero in zero field
because of symmetry, and only even derivatives have been analyzed \cite{muller,morita,morita2}
(in fact, equations similar to (\ref{ten}) and (\ref{twelve}) for the second and fourth
derivatives have been derived in
references \cite{morita,morita2} by neglecting odd correlations). 
However, this is true for any lattice in strictly zero field, as every spin configuration
has a mirror image (obtained by flipping all the spins) with exactly the same
energy, and inverted sign of the configurational magnetization $\sum_{i} S_i$.
The usual procedure of breaking the symmetry by retaining an infinitezimal
positive field while taking the thermodynamic limit, and only then taking the zero field limit,
is not suitable in the present case 
%\cite{morita}
because the derivatives of the free energy diverge 
in wide temperature regions (rather than just in a single critical point).
Here it seems more appropriate to break the symmetry by 
applying an infinitely strong local field of infinitezimal range,
which can be implemented by
restricting a single 
spin into one of the two possible orientations,
while considering all the possible orientations of all the other spins. 
In a recent work \cite{SSF3}, the present authors have
analyzed the effect of restricting a single spin in the $\{+\}$ orientation on magnetization
(first derivative of the free energy with respect to field),
for the current system. It was found that
symmetry is indeed broken by fixing an arbitrary (surface, bulk or center) spin. 
While without this restriction magnetization is identically zero in strictly zero field 
for arbitrary system size, 
it was shown 
%\cite{SSF3} 
that fixing any spin leads to magnetic ordering 
of extremely large systems, in a wide temperature range 
(even if magnetization does go to zero in the thermodynamic 
limit for all nonzero temperatures).

In the rest of this paper, we shall therefore adopt the strategy of breaking
the symmetry by fixing a single (central) spin in the $\{+\}$ (upward) orientation
while taking the thermodynamic limit, and we shall henceforth 
use the term $\ell$-th order susceptibility for the quantities
\begin{equation}
\chi_n^{(\ell)}\equiv\frac{\beta^{\ell-1}}{N_n}\,
\frac{1}{Z_{n}^{+}}
\left.{\frac{\partial^\ell Z_{n}^{+}}{\partial (\beta H)^\ell}}\right|_{H=0} 
\label{thirteen}
\end{equation}
for finite size systems, and
\begin{equation}
\chi^{(\ell)}\equiv\lim_{n\longrightarrow\infty} \chi_n^{\ell} ,
\label{fourteen}
\end{equation}
for the thermodynamic limit.

The restricted magnetization, $<m>^{+}\equiv\chi^{(1)}$ is found \cite{SSF,melin}
to be zero for all nonzero temperatures in the thermodynamic limit, 
(even if it retains nonzero values in a wide temperature range for systems 
far exceeding in number of particles the observable Universe \cite{SSF3}),
while 2-nd 
order susceptibility $\chi_n^{(2)}$ diverges below 
$T_2=2k_B^{-1}J\ln ({\sqrt 2}+1)$. 
It was also recently shown \cite{SSF2} that in the thermodynamic limit the 
divergence of $\chi^{(2)}$ in the vicinity and at $T_2$ is extremely weak, with 
critical exponent $\gamma=0$, where susceptibility of a finite tree $\chi_n^{(2)}$ 
diverges proportionally to the three generation level $n$, as $n\longrightarrow \infty$ . 
In the rest of this paper, we analyze the third and fourth order susceptibility
given by $\chi_n^{(3)}\equiv \beta^2{\cal U}_n/N_n$ and 
$\chi_n^{(4)}\equiv \beta^3{\cal V}_n/N_n$, respectively.

Equations (\ref{ten}-\ref{twelve}) are deceptive in the sense that 
at first glance they suggest divergence of second, third and fourth order susceptibilities
for arbitrary tree generation level $n$, 
at points $2 t^2=1$ and $2 t=1$ (corresponding to $T_2$ and $T_{BP}$, respectively),
while in addition, the fourth order susceptibility seems to diverge 
(irrespective of $n$) at $4 t^3=1$ and $8 t^4=1$ 
(the last corresponding to the M\" uller-Hartman and Zittartz temperature $T_4$ \cite{muller}).
In fact, the exact expression corresponding to (\ref{twelve}), but 
obtained by neglecting odd correlations \cite{morita},
indeed does diverge at $8 t^4=1$.
On the other hand, a more detailed analysis of 
(\ref{ten}-\ref{twelve})
shows that the interplay between the 
individual terms cancels out these aparent 
singularities at the mentioned temperature points. In particular, 
upon series expansion of the exact expressions (\ref{ten}-\ref{twelve}) arround
these points (from either side), it is found that there is no divergence for any finite $n$.
We give the leading terms of these expansions for large $n$ in Tabs.~\ref{tab1} and \ref{tab1b}.
\begin{table}
\caption{\label{tab1}Leading terms (for large $n$) of the series expansion
of susceptibilities $\chi_n^{(2)}$, $\chi_n^{(3)}$ and $\chi_n^{(4)}$, arround 
points $2t=1$ and $2t^2=1$, demonstrating that expressions (\ref{ten}-\ref{twelve})
do not diverge at these points for any finite $n$ (see text for details).}
\begin{ruledtabular}
\begin{tabular}{lll}
&$t\rightarrow 1/2$&$t\rightarrow 1/\sqrt {2}$\\
\hline
$\chi_n^{(2)}$ & $\left(9/4\right)\ln 3$  & 
$-n\left(\sqrt {2}+3/2\right)\ln \left(\sqrt {2}-1\right)$\\
$\chi_n^{(3)}$ & $n\left(27/8\right)\ln^2 3$  & 
$n\,{2}^{n/2}\left (21/\sqrt {2}+15\right )\left [\ln \left(\sqrt {2}-1\right)\right ]^{2}$\\
$\chi_n^{(4)}$ & $2^n\left(243/16\right)\ln^3 3$ & 
$-{n}^{2}{2}^{n}\left ({\frac {51}{2}}+18\,\sqrt {2}\right )\left [\ln \left(\sqrt {2}-1\right)\right ]^{3}$\\
\end{tabular}
\end{ruledtabular}
\end{table}
\begin{table}
\caption{\label{tab1b}Leading terms of $\chi_n^{(4)}$ arround $4t^3=1$ and $8t^4=1$. 
}
\begin{ruledtabular}
\begin{tabular}{ll}
&$\chi_n^{(4)}$\\
\hline
$t\rightarrow 2^{-\frac{2}{3}}$ &
${{2}^{n}}\left [\ln \left({\frac {{2}^{\frac{2}{3}}+1}{{2}^{\frac{2}{3}}-1}}\right)\right ]^{3}
\left ({\frac {417}{16}}\,{2}^{\frac{2}{3}}+{\frac {663}{16}}+33\;{2}^{\frac{1}{3}}\right )$\\
$t\rightarrow 2^{-\frac{3}{4}}$ &
${{2}^{n}}\left [\ln \left({\frac {{2}^{\frac{3}{4}}+1}{{2}^{\frac{3}{4}}-1}}\right)\right ]^{3}
\left (\frac{81}{8}\,{2}^{\frac{1}{2}}+\frac{33}{4}\,{2}^{\frac{3}{4}}+
{\frac {225}{16}}+12\;{2}^{\frac{1}{4}}\right )$
\end{tabular}
\end{ruledtabular}
\end{table}

Consequently, divergence of higher order susceptibilities in the thermodynamic limit
in different temperature regions is caused only by diverging generation level $n$, and the
finite size scaling is accomplished simply by the formula $\ln\chi_n^{\ell}/n$.
In Fig.~1 we show the scaled curves of higher order derivatives, obtained using
formulas (\ref{ten}-\ref{twelve}), for several system sizes.

It is seen from Fig.~1 that there are three distinct temperature regions, where
each temperature represents a point of second, third, or fourth order phase transition.
Between zero and $T_2$ the second derivative diverges (the first derivative being
finite), and we have a second order phase transition line. Between $T_2$
and $T_{BP}$ the third derivative diverges
(first and second derivatives being finite), 
representing a line of third order phase
transitions, while the fourth order derivative diverges between $T_{BP}$ and
$\infty$ (where all lower derivatives are finite), 
representing a fourth order phase transition line.
The explicit expressions for the limiting curves, corresponding to the
finite size system curves shown in Fig.~1,
are obtained by taking the limit 
$\kappa^\ell\equiv\lim_{n\longrightarrow\infty} {\ln\chi_n^\ell}/{n}$ ,
%\begin{eqnarray*}
%\kappa^\ell\equiv\lim_{n\longrightarrow\infty} \frac{\ln\chi_n^\ell}{n} ,
%\label{fifteen}
%\end{eqnarray*}
and the results obtained using formulas
(\ref{ten}-\ref{twelve}) for different temperature regions
are sistematized in Tab.~\ref{tab2}.

\begin{figure}
\includegraphics[width=2.7in]{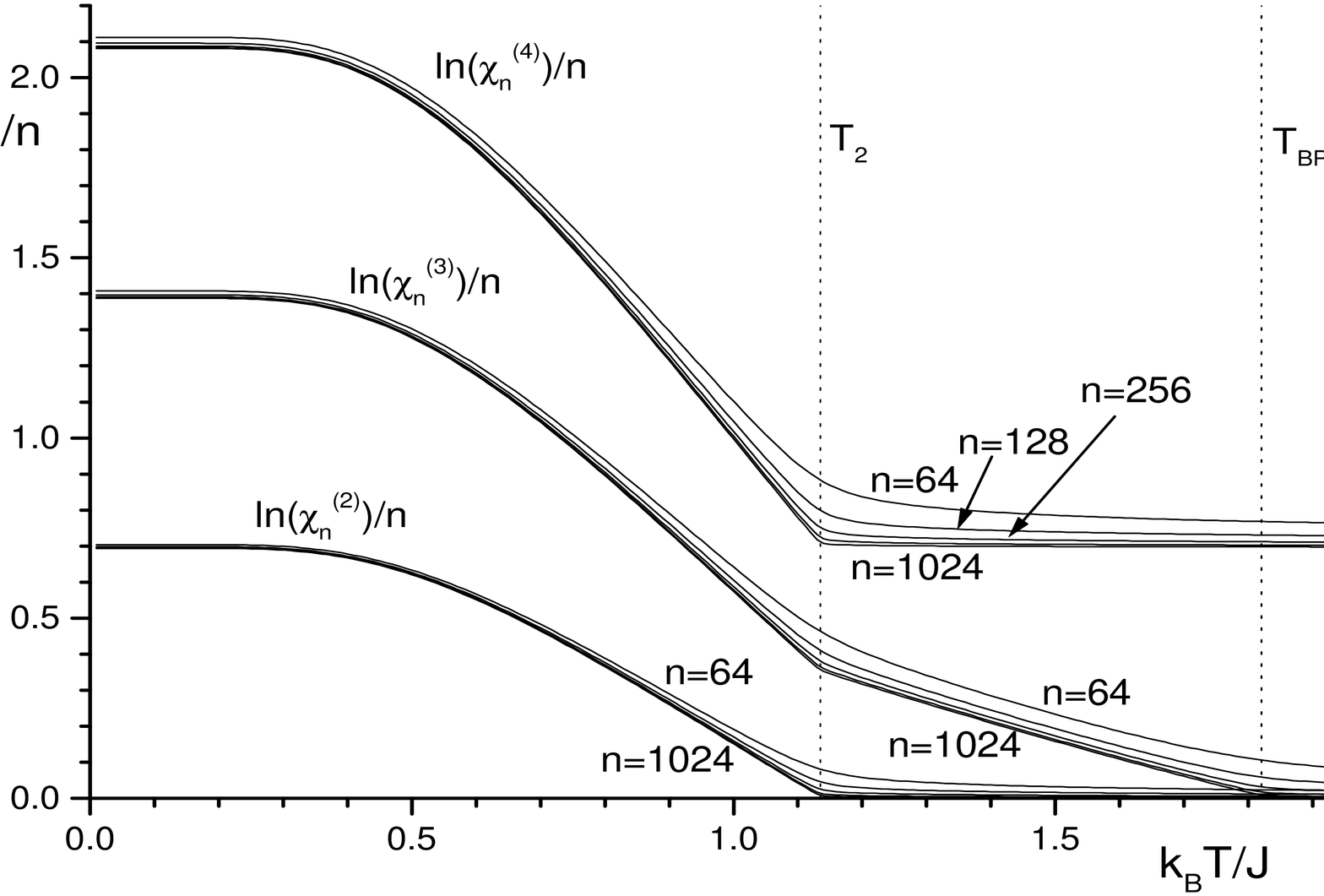}
\caption{\label{fig1} 
Scaled second, third and fourth-order zero-field susceptibilities, 
calculated using formulas (\ref{ten}-\ref{twelve}),
for several system sizes $n=64, 128, 256$ and $1024$.
The dotted vertical lines indicate the points $T_{2}$ and $T_{BP}$, where the
transition changes from second to third, and from third to fourth order, respectively.
}
\end{figure}

\begin{table}
\caption{\label{tab2}Scaled second, third and fourth order susceptibility,
in different temperature regions.}
\begin{ruledtabular}
\begin{tabular}{lccc}
&$0\leq T\leq T_2$ & $T_2\leq T\leq T_{BP}$ & $T\geq T_{BP}$\\
&$\left( 1\geq t\geq 1/\sqrt 2\right)$ & $\left( 1/\sqrt 2 \geq t \geq 1/2 \right)$ & $\left(t\leq1/2\right)$\\
\hline
$\kappa^{(2)}$ & $\ln\left(2\,t^2\right)$  & $0$  & $0$ \\
$\kappa^{(3)}$ & $\ln\left(4\,t^3\right)$  & $\ln\left(2\,t\right)$  & $0$ \\
$\kappa^{(4)}$ & $\ln\left(8\,t^4\right)$  & $\ln 2$  & $\ln 2$ \\
\end{tabular}
\end{ruledtabular}
\end{table}

In conclusion, after decades of continuous interest, the Ising model on the Cayley tree 
continues to furnish new insights into critical phenomena. Being a highly unphysical
system (with its infinite dimension and finite order of ramification), Cayley tree 
may turn out a unique structure where phase transitions of order higher then two 
indeed do exist, and the latter may prove to be of only academic interest.
On the other hand, it is possible that current findings may turn out relevant for 
interpretation of experimental data on finite size branching structures,
of the real observable physical world.


\begin{references}

{\bibitem{muller}
E. M\" uller-Hartmann and J. Zittartz, Phys. Rev. Lett. 33 (1974) 893.}

\bibitem{egarter}
T.P. Eggarter, Phys. Rev. B 9 (1974) 2989.

{\bibitem{heimburg}
J. von Heimburg and H. Thomas, J. Phys. C 7 (1974) 3433.}

{\bibitem{matsuda}
H. Matsuda, Prog. Theor. Phys. 51 (1974) 1053.}

{\bibitem{SSF}
T. Sto{\v s}i{\'c}, B.D. Sto{\v s}i{\'c} and I.P. Fittipaldi,
J. Mag. Mag. Mater. 177-181 (1998) 185.}

\bibitem{morita}
T. Morita and T. Horiguchi, Prog. Theor. Phys. 54 (1975) 982.

\bibitem{morita2}
T. Morita and T. Horiguchi, J. Stat. Phys. 26 (1981) 665.

\bibitem{melin}
R. M\' elin, J.C. Angl\' es d'Auriac, P. Chandra \\
and B. Dou\c cot, J. Phys. A 29 (1996) 5773.

\bibitem{SSF2}
T. Sto\v si\'c, B.D. Sto\v si\'c and I.P. Fittipaldi, 
Physica A, 320 (2003) 443.

\bibitem{SSF3}
T. Sto\v si\'c, B.D. Sto\v si\'c and I.P. Fittipaldi, 
cond-mat/0305581, to appear in Physica A (2005).

\end{references}
\end{document}